\documentclass[aps,showpacs,superscriptaddress,twocolumn]{revtex4}

\usepackage{amsfonts}
\usepackage{amsmath}
\usepackage{amsthm}
\usepackage{amscd}
\usepackage{amssymb}
\usepackage{subfigure}
\usepackage{amsxtra}
\usepackage{bm}           
\usepackage{bbm}
\usepackage{exscale}
\usepackage{epic,rotating}
\usepackage{graphics,color,dsfont}
\usepackage{latexsym}
\usepackage{enumerate}
\usepackage{dcolumn}      
\usepackage{pstricks,pst-grad,fancybox,graphics}
\usepackage[scanall]{psfrag}
\usepackage[dvips]{epsfig}

\newcommand{\R}{\ensuremath{\mathbbm R}}

\newcommand{\bra}[1]{\ensuremath{\langle#1|}}

\newcommand{\ket}[1]{\ensuremath{|#1\rangle}}

\newcommand{\braket}[2]{\ensuremath{\langle #1|#2\rangle}}

\newcommand{\ketbra}[1]{\ensuremath{| #1 \rangle \langle #1 |}}

\newcommand{\eins}{\ensuremath{\mathbbm 1}}
\newcommand{\HH}{\ensuremath{\mathcal{H}}}
\newcommand{\SCAL}{\ensuremath{\mathcal{S}}}
\newcommand{\NN}{\ensuremath{\mathcal{N}}}

\newcommand{\MM}{\ensuremath{\mathcal{M}}}

\newcommand{\bear}{\begin{eqnarray}}
\newcommand{\eear}{\end{eqnarray}}
\newcommand{\bearn}{\begin{eqnarray*}}
\newcommand{\eearn}{\end{eqnarray*}}
\newcommand{\kommentar}[1]{}

\newcommand{\mean}[1]{\ensuremath{\langle #1 \rangle}}
\newcommand{\proj}[1]{\ketbra{#1}}
\newcommand{\tr}[1]{\ensuremath{\mbox{Tr}\left( #1 \right)}}

\newcommand{\bc}{\begin{center}}
\newcommand{\ec}{\end{center}}
\newcommand{\rr}{\mathbbm{R}}

\renewcommand{\qed}{\ensuremath{\hfill \blacksquare}\medskip}

\newtheorem{thm}{Theorem}
\newtheorem{lem}[thm]{Lemma}
\newtheorem{corol}[thm]{Corollary}
\newtheorem{prop}[thm]{Proposition}
\newtheorem{defn}[thm]{Definition}
\renewcommand{\vr}{\ensuremath{\varrho}}
\def\bi#1\ei {\begin{itemize}#1\end{itemize}}
\def\bea#1\eea {\begin{align}#1\end{align}}
\def\bean#1\eean {\begin{align*}#1\end{align*}}
\def\ben#1\een {\begin{equation*}#1\end{equation*}}
\def\be#1\ee {\begin{equation}#1\end{equation}}
\def\bes#1\ees {\begin{equation}\begin{split}#1\end{split}\end{equation}}

\newcommand{\halbe}{\frac{1}{2}}

\begin{document}

\title{Multiparticle covariance matrices
and the impossibility of detecting graph state
entanglement with two-particle correlations}

\author{Oleg Gittsovich}

\affiliation{Institute for Quantum Computing,
University of Waterloo, 200 University Avenue West,
N2L 3G1 Waterloo, Ontario, Canada}

\affiliation{Institut f\"ur Quantenoptik und Quanteninformation,
~\"Osterreichische Akademie der Wissenschaften, Otto-Hittmair-Platz 1,
6020 Innsbruck, Austria}

\author{Philipp Hyllus}
\affiliation{INO-CNR BEC Center and Dipartimento di Fisica, Universit{\`a} di Trento, I-38123 Povo, Italy}

\author{Otfried G\"uhne}
\affiliation{Institut f\"ur Quantenoptik und Quanteninformation,
~\"Osterreichische Akademie der Wissenschaften, Otto-Hittmair-Platz 1,
6020 Innsbruck, Austria}

\affiliation{Institut f\"ur Theoretische Physik,
Universit\"at Innsbruck, Technikerstra{\ss}e 25,
6020 Innsbruck, Austria}

\begin{abstract}
We present a criterion for multiparticle entanglement based on covariance
matrices. On the one hand, the criterion allows to detect bound entangled
states which are not detected by other 
criteria; on the other hand,
some strongly entangled pure states such as the GHZ states
are not detected. We show, however, that this is a general
phenomenon: No separability criterion based on two-particle
correlations can recognize the entanglement in the family of
graph states, to which the GHZ states belong.
\end{abstract}


\date{\today}

\maketitle

\section{Introduction}
The presence of quantum correlations in physical states gives
several advantages in performing certain tasks. For instance,
entangled states
are used in quantum key distribution protocols, serve as a
resource in measurement based quantum computation and provide
insights into fundamental questions about quantum mechanics
\cite{nielsenchuang,reviews}.

Consequently, many efforts have been undertaken to characterize
entanglement. As a first step, one usually tries to prove inseparability
of a given state. In case the state is entangled, one may further
attempt to estimate the amount of the entanglement in the state. For
bipartite systems, entanglement criteria based on covariance matrices
have turned out to be powerful tools for detecting and quantifying
entanglement \cite{oldprl,wir, wirPRA, cmquanti}.

In this paper, we present an entanglement criterion based on covariance
matrices for multipartite entanglement. Our criterion is a generalization
of the covariance matrix criterion (CMC) derived in Ref.~\cite{wir} to the
multipartite setting. We show that the multipartite CMC is in some cases
a strong criterion, detecting bound entanglement in thermal states of spin
models, which is not detected by other commonly used criteria.
Surprisingly, some pure highly entangled states like Greenberger-Horne-Zeilinger
(GHZ) states are not detected by the multipartite CMC. However, this turns out
to be a quite general phenomenon coming  from the locality of the observables
used in the covariance matrices. In particular, we show that
all two-particle correlations of graph states (a family of highly entangled
multi-qubit states that is important for many applications \cite{reviewgs})
are compatible with a fully separable state. Consequently, {\em no} separability
criterion, which uses only the information gained from the two-party reduced
density matrices can detect graph state entanglement. This proves that not
only the multipartite CMC cannot detect graph states, but also
optimal spin squeezing inequalities \cite{spsqueez}, entanglement
witnesses based on two-particle Hamiltonians \cite{hamiltonian} or
structure factors \cite{KKBBKMprl09} or criteria based on magnetic
susceptibility measurements \cite{marcinnjp} will fail to do that.
A previously noted result in this direction is that the entanglement of
linear cluster states, which constitute a subset of graph states,
cannot be detected by spin squeezing inequalities \cite{gezacluster}.

Our paper is organized as follows. In Sec.~\ref{gencrit} we will first recall the basic
notions of covariance matrices and of the bipartite CMC. Then, we present the multipartite
generalization of the CMC and discuss its evaluation. For multi-qubit systems,
one can evaluate the criterion via semidefinite programming and for tripartite systems we
present an analytical approach. In Sec.~III we consider examples of states that are detected by the
multipartite CMC. We first consider pure three qubit states, and then bound entangled thermal states
in spin models. It turns out that of these states the multipartite CMC detects a larger
subset than the optimal
spin squeezing inequalities. In Sec.~IV we show that the two-particle correlations
in any connected graph state with more than two qubits are indistinguishable from the correlations
of some fully separable state. Finally, we conclude and discuss further research directions.

\section{General criterion and its evaluation}
\label{gencrit}
We begin this section by recalling the definition of the covariance
matrix (CM) for the bipartite case and stating the covariance
matrix criterion (CMC) for two parties. This will serve as a fundament
for the generalization on three parties, which can be easily generalized
to any number of parties.

\subsection{The bipartite CMC}
Let us start with the definition of CMs and some
basic facts about the CMC. A detailed description
can be found in Ref.~\cite{wirPRA}.

\begin{defn}
Let $\varrho$ be some quantum state acting on the Hilbert space
$\HH$ and let $\{M_k\}$ be a set of observables. Then, the
covariance matrix $\gamma$ of the state $\varrho$ is a real
and symmetric matrix, given by the entries
\be
\gamma_{kl}(\varrho)=\halbe\mean{M_k M_l + M_l M_k}_{\varrho}-\mean{M_k}_{\varrho}\mean{M_l}_\vr.
\label{cmgener}
\ee
\end{defn}
\noindent
In what follows we will write $\gamma_{kl}$ instead of
$\gamma_{kl}(\varrho)$, since it is usually clear which
quantum state is considered.

The CMC for the bipartite case formulated in Refs.~\cite{wir, wirPRA}
makes use of a block form of the CM. The block structure of the CM
arises naturally if one considers only local observables:

\begin{defn}
\label{cmbipadef} Let $\varrho=\varrho_{AB}$ be a bipartite quantum state defined on the Hilbert space
$\HH_A\otimes\HH_B$ of dimension $d=d_Ad_B$ and $\{A_k\otimes\eins,\eins\otimes
B_l\}$ a set of local observables. Then the covariance matrix of the state
$\varrho_{AB}$ has a block structure
\be
\gamma_{AB} = \left(\begin{array}{cc}A & C\\ C^{T}&B
\end{array}\right),
\label{cmbipa}
\ee
where $A$ and $B$ are CMs of the reduced states and the matrix
$C_{kl}=\mean{A_k\otimes B_l}-\mean{A_k}\mean{B_l}$ contains the correlations
between the subsystems. For product states $\vr=\vr_A \otimes \vr_B$ the matrix
C vanishes.
\end{defn}

Up to now, we did not specify the observables which are used in
our approach. In the following, we will always assume that the
observables $A_k$ (and similarly $B_k$) are an orthonormal basis
of the corresponding operator space. This means that they fulfill
$Tr(A_i A_j) = \delta_{ij}$; for example, if Alice has a qubit,
one can take
$\{A_i\} =\{\eins/\sqrt{2}, \sigma_x/\sqrt{2},\sigma_y/\sqrt{2}, \sigma_z/\sqrt{2}\}.$
The later results are, however, independent of {\it which} basis of observables
has been chosen.

Let us now formulate the general CMC, which has already been derived in
Ref.~\cite{wir}. This will make the generalization on the multipartite case
more clear.

\begin{prop}[Bipartite CMC]
\label{CMCbipartite}
Let $\gamma_{AB}$ be a block CM of a bipartite separable state $\varrho_{AB}$.
Then there exist pure states $\ket{\psi_k}$ in $\HH_A$ and $\ket{\phi_k}$ in
$\HH_B$ and probabilities $p_k$ such that if we define $\kappa_A=\sum_k p_k
\gamma(\ket{\psi_k}\bra{\psi_k})$ and $\kappa_B= \sum_k
p_k\gamma(\ket{\phi_k}\bra{\phi_k})$ the inequality
\be
\gamma_{AB}(\vr_{AB}) \geq\kappa_A\oplus\kappa_B
\Leftrightarrow
\begin{pmatrix} A & C\\
                C^T & B
\end{pmatrix}\geq
\begin{pmatrix} \kappa_A & 0 \\
                0        & \kappa_B
\end{pmatrix}
\label{CMC}
\ee
holds. This means that the difference between left and right hand side must be
positive semidefinite. If there are no such $\kappa_{A,B}$ then the state
$\rho$ must be entangled.
\end{prop}

One can extend this criterion, by looking for the largest $t$, such that
$\gamma_{AB}(\vr_{AB}) \geq t \kappa_A\oplus\kappa_B$ holds. Then, for
separable states this inequality is fulfilled for  $t = 1$, while for entangled
states it can happen that only smaller values are allowed. This parameter $t$
can then be used for a quantification of entanglement by giving lower bounds
on the concurrence, for details see Ref.~\cite{cmquanti}.

A central difficulty in this criterion is the characterization of the possible
$\kappa_{A,B}$: How is it possible to guarantee that no $\kappa_{A,B}$ exist? For this
problem, however, several results have been obtained. Consequently, the CMC
has been shown to be a strong criterion for bipartite entanglement: It
detects many bound entangled states, contains several known criteria
as corollaries, and, with the help of local filtering operations,
it is necessary and sufficient for two qubits \cite{wir, wirPRA}.

\subsection{The multipartite CMC}

Let us now start with the generalization of the CMC to multipartite systems.
We formulate our conditions for tripartite systems, but the generalization to
more parties is then straightforward. First, in analogy to Definition \ref{cmbipadef}
we consider a CM of a tripartite state $\varrho_{ABC}$ by using only local observables.
By doing so we arrive at the definition of the block CM for tripartite quantum states:

\begin{defn}
\label{cmtripadef}
Let $\varrho=\varrho_{ABC}$ be a tripartite quantum state defined on the Hilbert space
$\HH_A\otimes\HH_B\otimes\HH_C$ and $\{A_k\otimes\eins\otimes\eins,\eins\otimes
B_l\otimes\eins,\eins\otimes\eins\otimes C_m\}$ a set of local observables.
Then the covariance matrix of the state $\varrho_{ABC}$ has a block structure
\be
\gamma = \left(\begin{array}{ccc}A & D & E\\ D^{T}&B & F\\
E^{T}&F^{T}&C\end{array}\right).
\label{cmtripa}
\ee
where $A,B$, and $C$ are CMs of the reduced single particle states and
the matrices
$D_{kl}=\mean{A_k\otimes B_l}-\mean{A_k}\mean{B_l}$ etc.~contain
the correlations between the subsystems.
\end{defn}

Similar to the bipartite case, one can easily see that for any fully
product state $\vr=\vr_A \otimes \vr_B \otimes\vr_C $ the matrix $\gamma$
takes a
block-diagonal form. Hence, using the concavity property of
the CMs we arrive at the generalization of the CMC on the tripartite case:

\begin{thm}[Tripartite CM]\label{tripartCMC}
Let $\varrho$ be a fully separable tripartite state. Then there exist states
$\ket{\phi_\alpha^{(k)}}$ and matrices $\kappa_\alpha$ of the
form $\kappa_{\alpha}=\sum_k p_k\gamma(\ketbra{\phi_\alpha^{(k)}})$
for $\alpha = A, B, C$ such that
\be
\gamma  \geq  \left(\begin{array}{ccc}\kappa_A & 0 & 0\\ 0&\kappa_B & 0\\
0&0&\kappa_C\end{array}\right).
\label{cmc3p}
\ee
If no such matrices $\kappa_{\alpha}$ exist, the state is entangled.
\end{thm}
{Proof.} The proof is similar than for the bipartite CMC \cite{wir}.
Any fully separable state can be written as
\be
\vr= \sum_k p_k \ketbra{\phi_A^{(k)}} \otimes \ketbra{\phi_B^{(k)}} \otimes \ketbra{\phi_C^{(k)}}
\ee
where the $\{p_k\}$ form a probability distribution. Using now the concavity property
of the CMs, stating that
$\gamma (\sum_k p_k \vr^{(k)}) \geq \sum_k p_k \gamma (\vr^{(k)})$
and the fact that for the product states $\gamma$ is block diagonal proves the claim.
\qed

While the multipartite CMC is easy to prove, the complicated part is again its
evaluation: How can we exclude that the $\kappa_\alpha$ exist? In the next paragraphs
we present two strategies to do that. Firstly, for multi-qubit systems,
one can resort to
semidefinite programs (SDPs) and formulate the problem of searching over all
possible $\kappa_{\alpha}$ as a feasibility problem \cite{SDP}. Secondly, one
can deduce computeable criteria from the positivity semidefiniteness of the
matrix in Eq.~(\ref{cmc3p}).

\subsection{Evaluation of the multipartite CMC for qubits using SDP}
%

For qubits, the positivity test in Eq.~(\ref{cmc3p})
can be formulated in a simple way as a special instance
of an efficiently solvable SDP, namely as a feasibility
problem. This is true for any number of qubits, but
cannot be easily extended to higher dimensional systems.

In order to do so we first note that in the case of qubits
one can give an exact characterization of $\kappa_{\alpha}.$
Let us assume that the used observables are
$\{A_i, i=1,...,4\} = \{\eins/\sqrt{2}, \sigma_x/\sqrt{2},\sigma_y/\sqrt{2}, \sigma_z/\sqrt{2}\}.$
Then, as one can directly check, the first row and column
(corresponding to $A_1= \eins/\sqrt{2}$) of the single-qubit
CM vanishes. So the criteria can effectively be formulated
with $3\times 3$-matrices $\kappa.$ These matrices were completely
characterized in Ref.~\cite{wirPRA} in terms of vectors in $\rr^3$:

\begin{lem}
For any vector $\ket{\phi}\in \R^3$ a matrix of the form
$(\eins_3 -\ketbra{\phi})/2$, is a valid CM of some pure
single qubit state. Consequently, if Alice owns a qubit
the set of valid $\kappa_A$ is given by all matrices of the form
\be
\kappa_A  = \halbe (\eins_3-\rho_A),
\label{zweiqubitkappa}
\ee
where $\rho_A$ is a real $3\times 3$ matrix with trace one and positive
eigenvalues.
\end{lem}

Moreover the characterization of $\kappa$ for one qubit provided by the last
Lemma is exhaustive in the following sense:

\begin{lem}\label{lemchar}
Consider the CMC condition for three qubits, $\gamma \geq \kappa_A \oplus \kappa_B \oplus \kappa_C.$
Then, there exist positive matrices $X_A, X_B$ and $X_C$ such that $Tr(X_{A/B/C})=1$
and $\gamma \geq X_A \oplus X_B \oplus X_C$ if and only if there exist valid $\kappa_{A/B/C}$
that fulfill the CMC condition.
\end{lem}
{\it Proof:} The sufficiency is given by the last Lemma; clearly,
the $\kappa_{A/B/C}$ can be directly used as $X_{A/B/C}$.
On the other hand, suppose we have found such matrices
$X_{A/B/C}$. We know, that the $\kappa_i$s for qubits must
have a special form, namely $\kappa_i = \halbe\left(\eins_3 - \rho_i \right)$.
Since the $X_i$ satisfy the condition $\gamma
\geq X_A\oplus X_B\oplus X_C$ and $\gamma$ has eigenvalues less or equal than
$\halbe$ \cite{wirPRA}, the $X_i$ have also eigenvalues smaller than $\halbe$,
and any $X_i$ can be written in the form $X_i = \halbe\left(\eins_3 -
x_i\right)$, where $Tr(x_i)=1$ and $x_i\geq 0$.
Therefore, $X_i$ is exactly of the form
of Eq.~(\ref{zweiqubitkappa}).
\qed

Using the last two Lemmata we can formulate the CMC for the
three-qubit case in the following form:
\begin{align}
\mbox{maximize: } & t
\nonumber\\
\mbox{subject to: }  &\gamma  -
tX_A\oplus X_B\oplus X_C\geq 0,
\nonumber\\
&
X_{A,B,C}\geq 0,
\nonumber\\
&
Tr(X_{A,B,C})=1.
\label{sdph}
\end{align}
If the maximal $t$ is larger or equal one, then the state is not detected
by the CMC, but if the above maximization has a solution only for $t<1$
then the state is detected by the CMC and is therefore entangled.

The problem in Eq.~(\ref{sdph}) can be readily re-written as a semidefinite program (SDP).
Such a problem can directly be solved using packages like SEDUMI or YALMIP \cite{SDP}.
The above formulation is a simplification of a similar formulation for
two qubits derived in Ref.~\cite{wirPRA}. Note that the solution of
Eq.~(\ref{sdph}) directly allows to determine the parameter $t$, so
the result might be used  for further quantification. Finally, note that
the SDP can be directly extended to test the multipartite CMC for more
than three qubits. Furthermore, if higher dimensional systems are considered,
one can also test the CMC via a SDP (since the $\kappa_{i}$ are always
nonnegative and have a fixed trace \cite{wirPRA}), however, this is not an
exhaustive characterization of the $\kappa_{i}$ and the SDP is therefore
weaker than the original CMC.

The multipartite criterion evaluated by the SDP detects many interesting bound
entangled states, which examples will be discussed in the next section. For
the time being we discuss a method of analytical evaluation of the tripartite CMC.

\subsection{Analytic test of non-separability of tripartite states}

In contrast to the previous approach, the analytical method works only
for the tripartite case, but it is not restricted to qubits.

What are the possible corollaries of the non-negativity of the block
matrix in Eq. (\ref{cmc3p})? First of all, we know that all submatrices
of this matrix have to be non-negative. As we learn from the bipartite case
\cite{wirPRA}, the strategy of considering submatrices leads to a quite
strong and easily computable separability criterion. In order to investigate
whether a tripartite state is fully separable or not, it is then natural
to consider $3\times 3$ submatrices of the CM in the block form.
To do this, we need the following fact which is true for any real $3\times 3$
non-negative matrix:

\begin{lem}
\label{lemma3part}
Let $\eta$ be a $3\times 3$ real positive matrix
\be
\eta=\left(\begin{array}{ccc} a & d & e\\ d & b & f\\
e & f & c\end{array}\right)\geq 0,
\label{3x3pos}
\ee
then
\be
\tilde{\eta}=
\left(\begin{array}{ccc} a & |d| & |e|\\ |d| & b & |f|\\
|e| & |f| & c\end{array}\right)\geq 0
\ee
is also a positive matrix.
\end{lem}
\noindent
This Lemma is an example of a norm compression inequality and the proof
is given in the Appendix.

Using this property of $3\times 3$ non-negative matrices
we can prove a proposition, which gives us an easily computable
separability test for tripartite states:
\begin{prop}\label{prop3part}
If the state $\varrho$ on a $d\times d\times d$-system is separable, then
\be
\left(\begin{array}{ccc}Tr(A)-d+1&\sum_i|d_{ii}|&\sum_i|e_{ii}|\\
                        \sum_i|d_{ii}|&\!\!\!\!\!Tr(B)-d+1&\sum_i|f_{ii}|\\
                        \sum_i|e_{ii}|&\sum_i|f_{ii}|&\!\!\!\!\!Tr(C)-d+1
      \end{array}\right)\geq 0.
\label{prop9}
\ee
Here, $d_{ij}, e_{ij}, f_{ij},$ are the elements of the blocks $D,E,F$ in Eq.~(\ref{cmtripa}),
respectively.
\end{prop}
{\it Proof.} Since $\gamma-\kappa_A \otimes \kappa_B \otimes \kappa_C$ is nonnegative,
all $3 \times 3$ submatrices are positive. So we consider the matrices
\be
\left(
\begin{array}{ccc} a_{ii}-(\kappa_A)_{ii} & d_{ii} & e_{ii}\\ d_{ii} & b_{ii}-(\kappa_B)_{ii} & f_{ii}\\
e_{ii} & f_{ii} & c_{ii}-(\kappa_C)_{ii}\end{array}\right)\geq 0.
\ee
Applying now Lemma \ref{lemma3part} and summing over all $i$ gives the claim,
if we use in addition the fact that $Tr(\kappa_{\alpha})=d-1$ for $\alpha=A,B,C$,
shown in Ref.~\cite{wirPRA}.
\qed

\section{Examples}
In this section we consider some examples for the previously derived
criteria. We will first consider random pure three-qubit states, and
then thermal states in spin models, which may be bound entangled.

\subsection{Random three-qubit states}
\label{randomstates}
In order to test the performance of the derived entanglement
criteria we first investigate three qubit pure states. According to Ref.~\cite{acinetal2000prl}
any three-qubit state can be brought by local unitary transformations  into a
{generalized Schmidt form},
\begin{align}
\ket{\psi}
&= \lambda_0\ket{000} + \lambda_1 \mbox{e}^{i\phi} \ket{100}
+ \lambda_2\ket{101} + \lambda_3\ket{110} + \lambda_4\ket{111},
\nonumber\\
& \mbox{ with } \lambda_i\geq 0, \: 0 \leq\phi\leq\pi, \: \sum_i\lambda_i^2=1.
\end{align}
Varying the real parameters $\lambda_i$ and $\phi$ one obtains
different families of entangled states.

Proposition \ref{prop3part} detects almost all of these entangled three-qubit
states: {Choosing the Pauli matrices as observables and the parameters
$\lambda_i$ is uniformly distributed between 0 and 1, whereas $\theta$ uniformly
distributed between 0 and $\pi$, one is able to detect 100$\%$ of the randomly
generated states.} However, Proposition \ref{prop3part}
fails to detect entanglement is the case of GHZ states,
$\ket{{\rm GHZ}}=(\ket{000}+\ket{111})/\sqrt{2}.$
Nontheless, even a small perturbation of the GHZ state avoids this problem, e.g.,
the state
\be
\ket{\psi^{\prime}_{{\rm GHZ}}} = \frac{1}{\NN}\left(\ket{000}+10^{-5}\ket{110}+\ket{111}\right)
\ee
is detected, here $\NN$ denotes the normalization.

In order to test whether this is an issue of the Proposition \ref{prop3part}
or whether is a property of the general CMC as stated in Theorem \ref{tripartCMC}
we used the first strategy and evaluated the general criterion with an SDP. The SDP
detects also 100 \% of the randomly generated states (in both cases $10^4$ instances were tested),
but it also fails to detect GHZ states. 

This phenomenon can be explained as follows: In the CMC as in Theorem \ref{tripartCMC}
uses only one- and two-point correlations (i.e. expectation values like $\mean{\sigma_x^A}$
or $\mean{\sigma_x^A \sigma_x^B}$) but never higher order correlations (like three-point
correlations $\mean{\sigma_x^A \sigma_x^B \sigma_x^C}$), as only the former appear in the
block structure of the CM.
However, considering only one and two-point correlations, it is impossible to
distinguish $\ketbra{{\rm GHZ}}$ from fully separable state $(\ketbra{000}+\ketbra{111})/2$,
as for both states the reduced one- and two-particle density matrices are the same.

As we will see in Sec.~\ref{graphstates} this is not an exceptional case.
{\it Any} state in the large class of graph states, which also contains the GHZ states,
cannot be distinguished from a fully separable state by considering one- and two-particle
correlations. This does not only affect the multipartite CMC, but it also means
that these states can never be detected by other approaches like spin squeezing
inequalities \cite{spsqueez}, or witnesses based on two-body Hamiltonians
\cite{hamiltonian} or structure factors \cite{KKBBKMprl09}.

\subsection{Entanglement in thermal states}
\label{thermstatessec}
Now we discuss thermal states of spin models and investigate their
entanglement properties using the approach of Eq.~(\ref{sdph}) and
Proposition \ref{prop3part}.

In Ref.~\cite{christianda} it has been shown that thermal states of spin
models with a small number of spins can have interesting separability properties.
This has been investigated frequently with spin squeezing inequalities
\cite{spsqueez,spsqueezpra}. They can be formulated as follows. First, one
takes the collective spin operators $J_k = \halbe\sum_{i=1}^N \sigma_{\vec n_k}^{(i)}$
where the vectors $\{{\vec n}_k\}_k$ are orthonormal.
Then one defines the matrices
\begin{eqnarray}
C_{kl} &=&
\frac{1}{2}\mean{J_k J_l + J_l J_k},
\nonumber
\\
\Gamma_{kl} &=& C_{kl} - \mean{J_k}\mean{J_l} \vphantom{\frac{1}{2}},
\nonumber\\
\chi &= &(N-1)\Gamma + C, \vphantom{\frac{1}{2}}
\end{eqnarray}
Given these matrices, a separable state fulfills \cite{spsqueezpra}:
\bea
\tr{\Gamma}&\geq\frac{N}{2},
\label{sseq1}
\\
\lambda_{\rm min}(\chi)&\geq \tr{C} - \frac{N}{2},
\label{spinsqueezineq}
\\
\lambda_{\rm max}(\chi)&\leq (N-1)\tr{\Gamma} - \frac{N(N-2)}{4},
\label{sseq3}
\eea
where $N$ is the number of spin-$\halbe$ particles. If one of these inequalities is violated,
the state must be entangled. Moreover, if only the {\it averaged} two-particle correlations
are known, then in many situations these inequalities are the strongest entanglement criteria
possible, in the sense that for any state that fulfills these criteria, there is indeed a separable
state with the same $C, \Gamma$ and $\chi$ \cite{spsqueezpra}.

The thermal states, which were considered in Refs.~\cite{christianda, spsqueez,spsqueezpra},
are the thermal states of the Hamiltonian
\be
H=\vec{S}_1\cdot \vec{S}_2+ \vec{S}_2\cdot\vec{S}_3 +
\vec{S}_1\cdot\vec{S}_3 +
h\left(\sigma_z^1+\sigma_z^2+\sigma_z^3\right).
\label{heisbham}
\ee
where $\vec{S}_k=(\sigma_x^{(k)},\sigma_y^{(k)},\sigma_z^{(k)}).$
This Hamiltonian describes three spin-$\halbe$ particles interacting
via the Heisenberg interaction and subjected to an external homogeneous
magnetic field of the strength $h$. The thermal states
\be
\varrho_T(h)=\frac{\mbox{e}^{-H/T}}{\tr{\mbox{e}^{-H/T}}}
\ee
represent a two-parametric family of density matrices. Using the
spin-squeezing inequalities one can find regions in the \emph{T-h}
diagram where all thermal states have a positive partial transposition
(PPT) with respect to any bipartition (and therefore do not violate
the PPT criterion \cite{PPTperes1996prl}), but the spin squeezing inequalities
show that the state is not fully separable. For $h=0$  and certain temperatures
the thermal state is even separable with respect to any bipartition, but
not fully separable \cite{spsqueez,spsqueezpra}.

As a first example, we can test our general criterion Eq.~(\ref{cmc3p})
via semidefinite programming for these thermal states. It turns out, that
the multipartite CMC detects exactly the same amount of the states as the
spin squeezing inequalities. This can be understood as follows: On the one hand when the
spin-squeezing inequalities are applied to spin models, then usually Eq.~(\ref{sseq1})
detects the entanglement. This inequality, however, can be seen as derived from local
uncertainty relations for special observables \cite{marcinnjp, spsqueez}. On the
other hand, the bipartite CMC is known to be equivalent to local uncertainty relations
with arbitrary observables \cite{wir} and this connection clearly holds for the multipartite
case. So it is not surprising that for this spin model, the CMC is not weaker than the
spin squeezing inequality.

However, slightly changing the Hamiltonian (\ref{heisbham}) gives rise to yet another
family of the states with two parameters, for which the comparison of three criteria
(PPT, Spin Squeezing and CMC) shows that the CMC criterion can detect some states that
are not detected by the PPT criterion or the spin squeezing criteria.
This change of the Hamiltonian consists in changing of the direction of one of the
local magnetic fields, therefore destroying the spatial symmetry of the Hamiltonian
\be
H'=\vec{S}_1\cdot \vec{S}_2+ \vec{S}_2\cdot\vec{S}_3 +
\vec{S}_1\cdot\vec{S}_3 +
h\left(\sigma_z^1+\sigma_x^2+\sigma_z^3\right).
\label{heisZXZ}
\ee

\begin{figure}[t!]
\includegraphics[width=0.8\columnwidth]{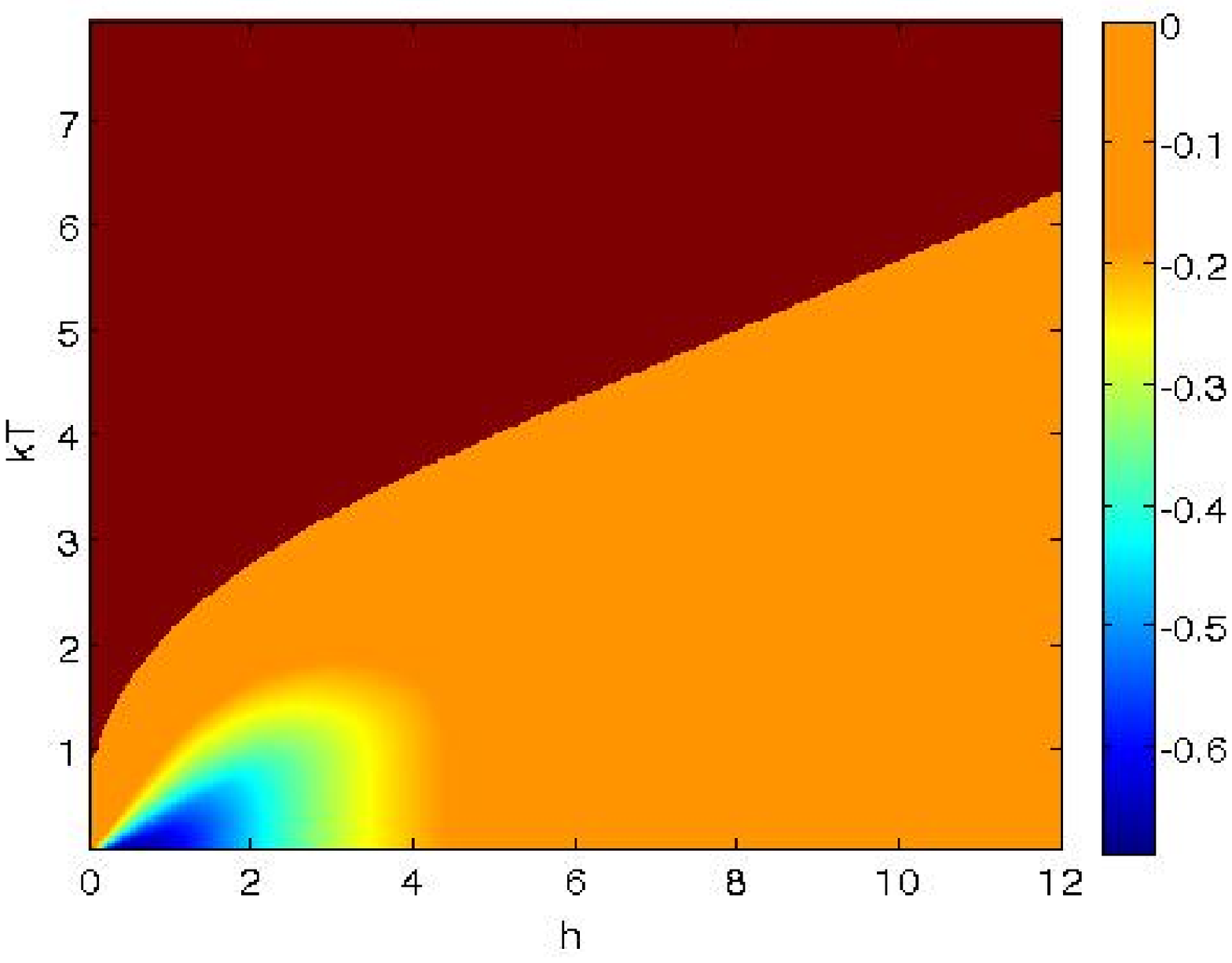}

(a)

\includegraphics[width=0.8\columnwidth]{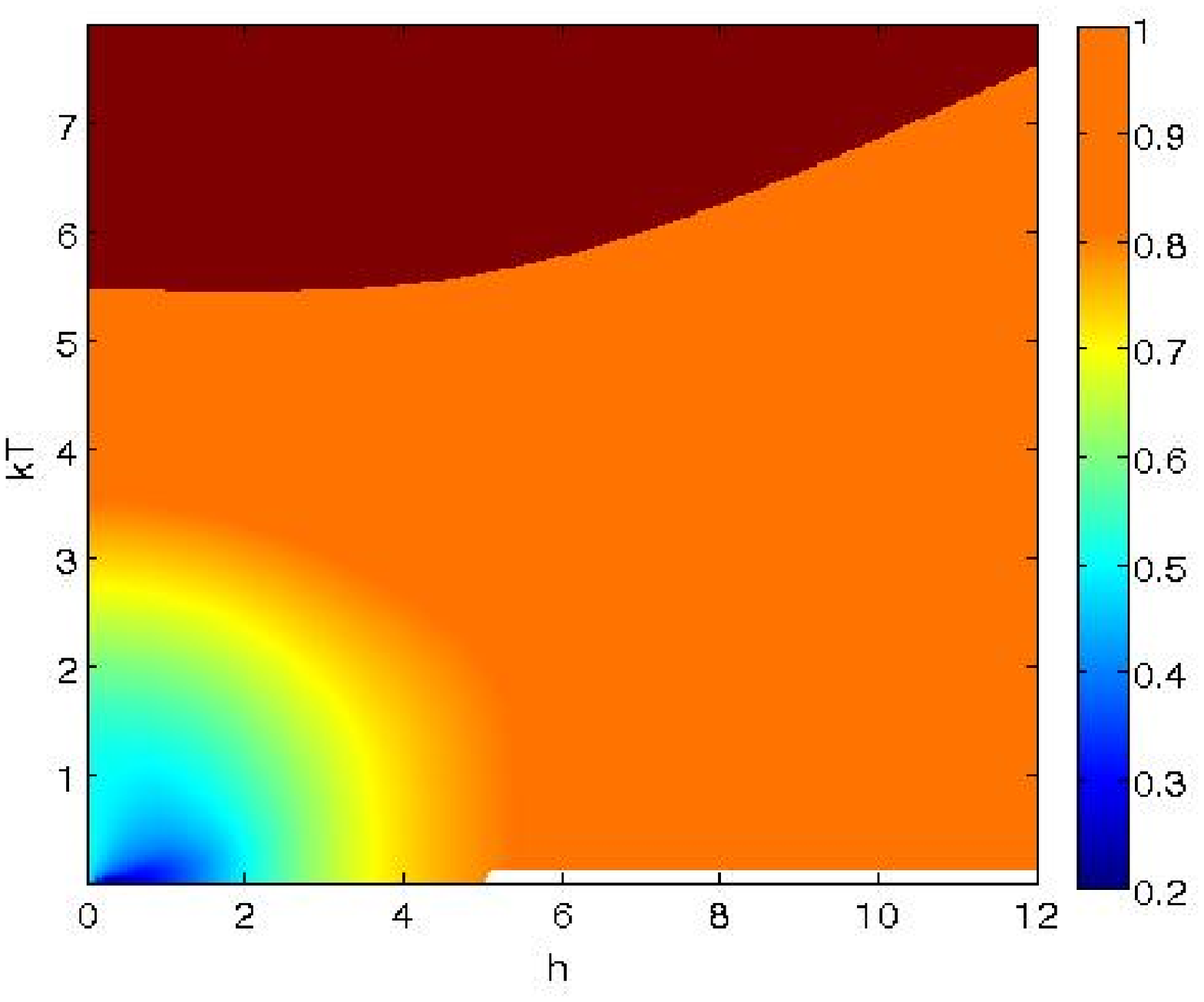}

(b)

\caption{(Color online) Detection of the thermal states of the Hamiltonian (\ref{heisZXZ})
for different values of temperature and external magnetic field by the multipartite
CMC. The dark red region corresponds to the states, which are not detected by
the criterion. (a) The criterion of Proposition \ref{prop3part} with the Pauli
matrices as observables. Different colors
correspond to the most negative eigenvalue of the matrix in Eq. (\ref{prop9}).
(b) Evaluation with the semidefinite program in Eq.~(\ref{sdph}). Here, different
colors correspond to different values of $t$.
\label{prop9plot}}

\end{figure}

\begin{figure}[t!]
\includegraphics[width=0.8\columnwidth]{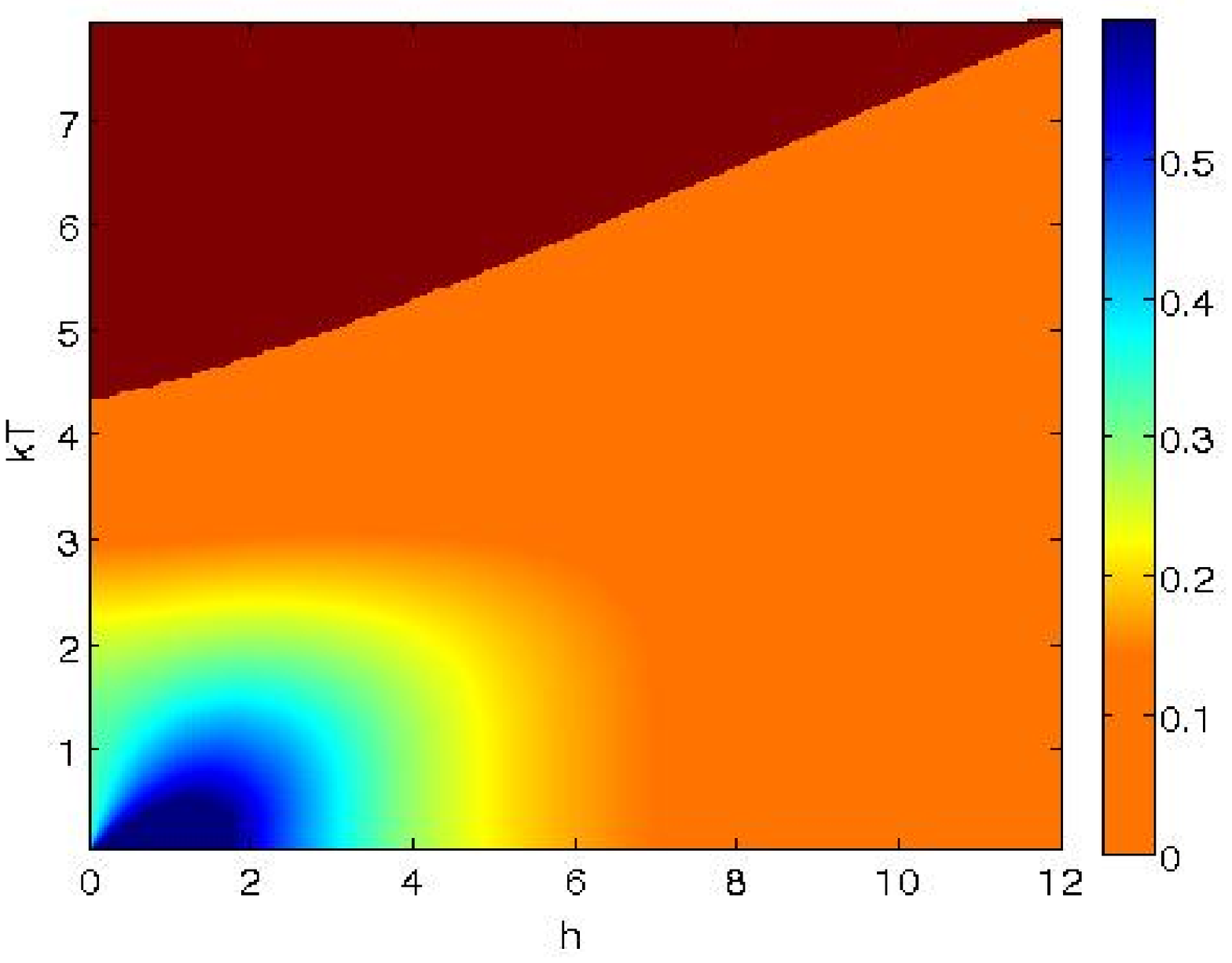}

(a)

\includegraphics[width=0.8\columnwidth]{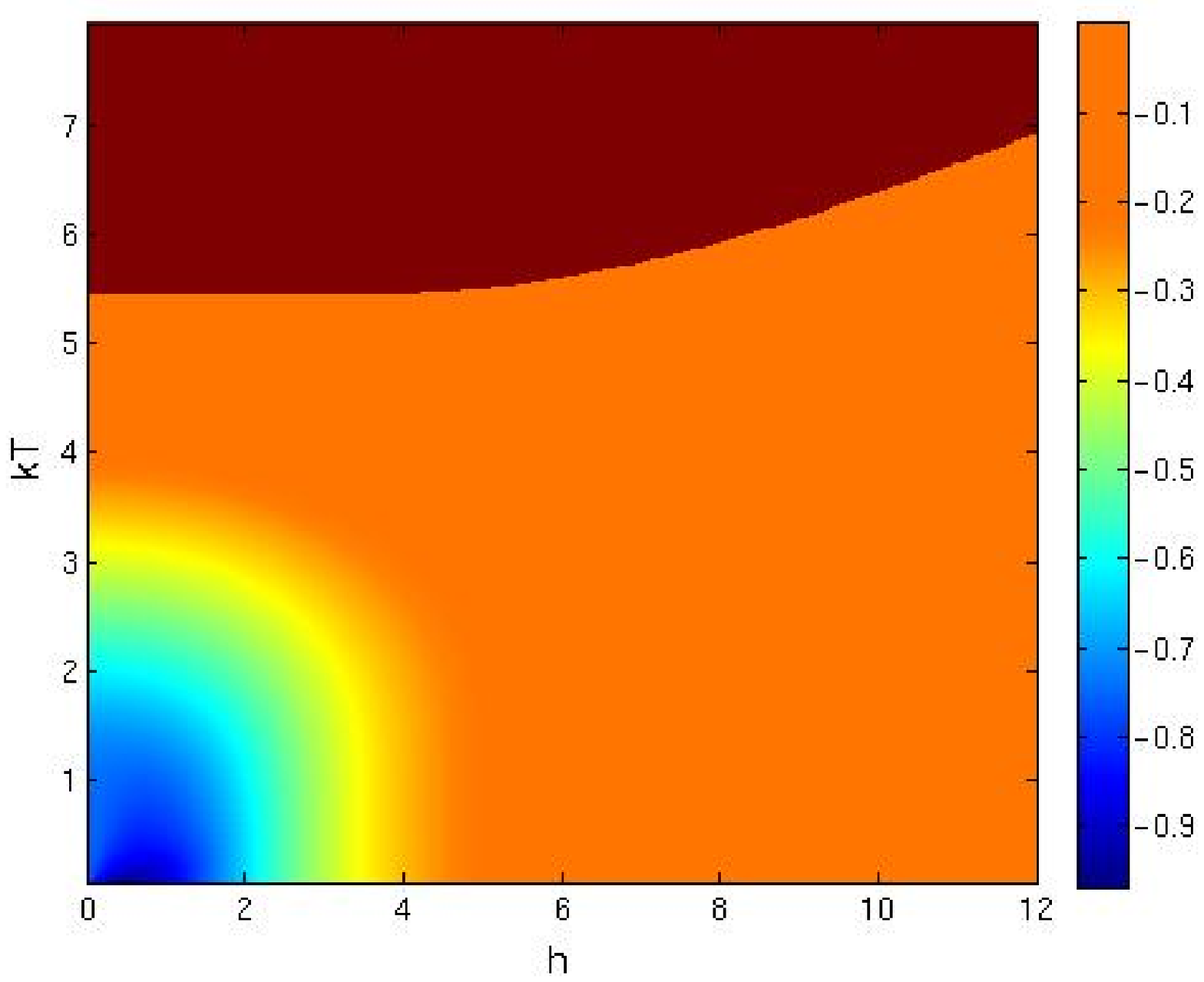}

(b)

\caption{(Color online) Detection of the thermal states of the Hamiltonian (\ref{heisZXZ})
for different values of temperature and external magnetic field by the PPT criterion
and the spin-squeezing inequalities. The dark red region corresponds to the states,
which are not detected by these  criteria. (a) The violation of the PPT
criterion characterized by the negativity \cite{vidalwernernegativity2002pra}. (b) The
detection of the states by spin squeezing inequalities (\ref{sseq1})-(\ref{sseq3}).
{Different colors  correspond to the maximal violation of the inequalities  (\ref{sseq1})-(\ref{sseq3}). Comparing this with Fig.~1(b) for values $kT \sim 7$, $h\sim 12$ we immediately see that the CMC for three qubits detects more states than the spin squeezing inequalities. These states violate, however, the PPT criterion as well. One finds an example of a bound entangled state, which is not detected by spin squeezing inequalities, but is detected by the CMC for the following values of parameters $kT = 5.533$, $h= 4.3$.}
\label{ZXZ}}
\end{figure}

Detection of these states by the CMC via Proposition \ref{prop3part} or
the SDP in Eq.~(\ref{sdph}) is presented in Fig.~\ref{prop9plot}. As expected,
the Proposition \ref{prop3part} is weaker than the SDP.
The detection of the PPT criterion and the spin-squeezing inequalities is
presented in Fig.~\ref{ZXZ}. Part (a) corresponds to the PPT criterion.
Here we characterize states by taking the maximum of three values of the negativity
corresponding to three different bipartitions
\cite{vidalwernernegativity2002pra}. Part (b) corresponds to the
detection of these states by the spin squeezing inequalities.  Comparing
Fig. 2(a) and Fig. 2(b) one sees that spin squeezing inequalities
are able to detect bound entanglement. More importantly, however, by comparing
Fig. 1(b) with Fig. 2(b) one sees that the CMC detects more states than the spin squeezing
inequalities, which one sees looking at the region of the plots where $kT \sim 7$, $h\sim 12$.
However, the CMC detects better not only bound entangled state, an example of a bound entangled
state, which is not detected by spin squeezing inequalities but violates the CMC is a thermal
state with parameters $kT = 5.533$, $h= 4.3$.
Note that this is not a contradiction to the optimality of the spin squeezing
criterion for some cases, since the CMC does not only use the
{averaged} two-point correlations.

\section{Entanglement in graph states and two-particle correlations}
\label{graphstates}
In Section \ref{randomstates} we have seen that the multipartite CMC
is unable to detect the entanglement in a three-qubit GHZ state. The
reason was that the CMC involves only two-particle correlations and
that the GHZ state $(\ket{000}+\ket{111})/\sqrt{2}$
and the fully separable state $(\ketbra{000}+\ketbra{111})/{2}$
have the same reduced two-particle density matrices, hence the
two-particle correlations are the same.

In this section we will show that this observation is not
a coincidence. We will consider the family of graph states
(defined precisely below) which comprises not only the GHZ
states, but also other states like cluster states, and which
is of eminent importance for many applications and experiments
\cite{reviewgs}.
We will show that for {\it any} connected graph state of three or more qubits
there is a fully separable state, which has the same one- and
two-qubit reduced density matrices. Hence, graph state entanglement
can never be detected by measuring two-particle correlations only,
and this is not only a restriction to the multipartite CMC, but also
to a variety of other entanglement criteria.

More precisely, we will prove:

\begin{thm}
Let $\ket{G}$ be a connected graph state with more than
two qubits. Then, there exists a fully separable
state $\varrho_G$ such that for any pair of
qubits $\{i,j\}$ the corresponding two-qubit
reduced density matrices of $\proj{G}$ and $\varrho_G$
coincide,
\be
\varrho_{ij}(\proj{G})=\varrho_{ij}(\varrho_G).
\label{claim}
\ee
Moreover, $\varrho_G$ has also the same one-particle reduced
density matrices, $\varrho_{i}(\proj{G})=\varrho_{i}(\varrho_G).$
\end{thm}

Here and in the following, we use the definitions
$\varrho_i(A)={Tr}_{I\backslash i}(A)$ and
$\varrho_{ij}(A)={Tr}_{I\backslash i,j}(A)$,
where $A$ is an operator acting on the Hilbert space of all $N$ qubits,
and $I$ is the set all particle indices. In a slight abuse
of the usual notation, we use the same definition below
even in situations where $A$, and hence $\varrho_i$ and $\varrho_{ij}$, are no quantum states.

The point of this theorem is that many entanglement criteria use only
two-particle correlations for the entanglement detection. Consequently,
these criteria must fail to recognize the entanglement in graph states:

\begin{corol}
Consider a graph state of three or more qubits. Then, the entanglement
cannot be detected using the following criteria:
\\
(a) the multipartite CMC from Eq.~(\ref{cmc3p}),
\\
(b) the optimal spin squeezing inequalities \cite{spsqueez},
\\
(c) entanglement witnesses based on two-body Hamiltonians \cite{hamiltonian},
and
\\
(d) entanglement witnesses based on structure factors \cite{KKBBKMprl09} or inequalities
using the magnetic susceptibility \cite{marcinnjp}.
\end{corol}

Before proving the Theorem, let us give a short definition
of graph states, for a broader discussion see Ref.~\cite{reviewgs}.
Consider a graph with $N$ vertices and some edges connecting
them. We are only interested in connected graphs which do
not separate into two unconnected parts. For any vertex $i$ we can
write down an operator
\be
g_i = X_i \bigotimes_{j\in \NN(i)} Z_j,
\ee
where here and in the following, $X_i, Y_i,$ and $Z_i$ denote the Pauli
matrices acting on the $i$-th qubit and $\NN(i)$ is the neighborhood
of $i,$ that is, all vertices that are connected with $i$. The graph state $\ket{G}$ is now
defined as the unique $N$-qubit eigenstate of all the $g_i,$
\be
\ket{G}=g_i \ket{G}.
\ee
The graph state is not only an eigenstate of the $g_i,$ but
also of their products. All products of the $g_i$ form a
commutative group with $2^N$ elements, the so-called stabilizer
$\SCAL(G)$. Using the elements of the stabilizer, we can also write
the graph state as
\be
\ketbra{G}=\frac{1}{2^N}\sum_{S_k \in \SCAL(G)} S_k.
\label{gproj}
\ee
This relation will be useful in our proof, which can be
split into three parts: Firstly, we will distinguish the different
cases where $\vr_{ij}$ is not the identity. Secondly, we will
investigate in detail what kind of reduced states $\vr_{ij}(\ketbra{G})$
can occur. Finally, we will see that there exists a measurement of the type
$\MM = \bigotimes_{i=1}^N O^{(i)}$ with $O^{(i)} \in \{X_i, Y_i, Z_i\}$
which allows to compute all two-particle correlations of the state $\ket{\rm G}$
and consequently
all $\vr_{i}(\ketbra{G})$ and $\vr_{ij}(\ketbra{G}).$ For such a measurement,
however, there is always a separable state giving the same results as
the graph state.

{\it Part I ---}
Let us start by characterizing all one- and two-particle
reduced density matrices. First, note that for a connected
graph one always has $\vr_{i}(\ketbra{G})=\eins/2.$ This
can be seen from Eq.~(\ref{gproj}). For most of the
$S_k \in \SCAL(G)$ (which are products of the $g_i$)
we have $\vr_{i}(S_k) = 0$ since they act non-trivially
on qubits besides $i.$ Since $S_0\equiv\eins \in \SCAL(G)$
we have for this element that $\vr_{i}(S_0) = 2^{N-1} \eins.$
Furthermore, we can only have $\vr_{i}(S_k) \neq 0$
and $\vr_{i}(S_k) \neq 2^{N-1} \eins$ if there is a correlation operator with
the property $g_{i} = X_{i}.$ This, however, cannot occur
in a connected graph.

\begin{figure}[t!]
\includegraphics[width=0.8\columnwidth]{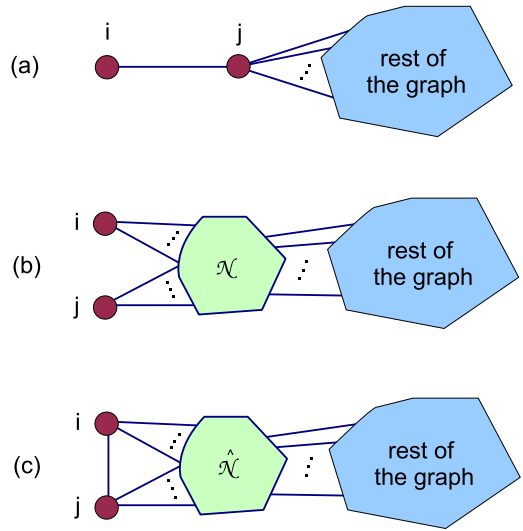}
\caption{(Color online) The three different cases where the reduced two-qubit density matrix
differs from the identity. In case (a) the qubit $i$ is connected to the qubit $j$ only,
and the qubit $j$ is connected to the rest of the graph R.  In case (b) the qubits
$i$ and $j$ are disconnected but have the same neighborhood $\NN(i)=\NN(j).$
In case (c) the qubits $i$ and $j$ are connected and have the same neighborhood apart
from $i,j$, {\em i.e.}, $\hat{\NN}=\NN(i) \setminus \{ j \} = \NN(j) \setminus \{ i\}$.
Small dots denote possible multiple connections.
See the text for further details.
\label{3cases}}
\end{figure}

Let us now turn to the reduced two-qubit density matrices.
As above, we find for most of the stabilizing operators
that $\vr_{ij}(S_k) = 0.$ Consequently for most of the qubit
pairs $\{ij\}$ we have that $\varrho_{ij}(\proj{G})$ is
completely mixed, {\em i.e.}, $\varrho_{ij}(\proj{G})=\eins/4$.
However, as already discussed in Ref.~\cite{notall},
there are three exceptions, see also Fig.~\ref{3cases}.
These exceptions occur if some $S_k$ acts non-trivially
only on two qubits $i$ and $j$. For these pairs
$\{ij\}$ there are three possibilities:

(a) Qubit $i$ has only one neighbor $j$ [$\NN(i)=j$],
which is coupled to the rest of the graph. Note that
there is no symmetry in interchanging the qubits $i$
and $j$. Then we have
\be
\varrho_{ij}(\proj{G}) = \frac{1}{4}(\eins + X_iZ_j).
\label{casea}
\ee

(b) The qubits $i$ and $j$ are not connected but have the
same neighborhood [$\NN(i)=\NN(j)$], which is connected to
the rest of the graph. This case is symmetric in $i$ and $j.$
The reduced state is
\be
\varrho_{ij}(\proj{G}) = \frac{1}{4}(\eins + X_i X_j).
\label{caseb}
\ee

(c) The qubits $i$ and $j$ are connected and have the
same neighborhood otherwise
which is connected to the rest of the graph. Defining
$\hat{\NN}(i)=\NN(i)\setminus \{j\}$ and
$\hat{\NN}(j)=\NN(j)\setminus \{i\}$
this means that
$\hat{\NN}(i) = \hat{\NN}(j).$
This case is also permutation symmetric.
We arrive at
\be
\varrho_{ij}(\proj{G}) = \frac{1}{4}(\eins + Y_iY_j).
\label{casec}
\ee

{\it Part II ---}
Let us now discuss the relations between the three cases above. Especially,
we would like to investigate in which cases one particle can contribute
to two or more cases.

\begin{itemize}

\item[(i)]
For a given pair of qubits $\{i_0,j_0\}$ maximally
one of the three cases can
apply.
\\
{\it Proof.}
Clearly, the case (b) excludes the case (c)
and vice versa, since in (c) the qubits are
connected and in (b) not. The same holds for
(a) and (b). If (a) and (c) hold
simultaneously, then $\hat{\NN}(i)=\hat{\NN}(j)=\emptyset$
which means that the graph consist only of two qubits
or is not connected. Both cases were excluded.

\item[(ii)]
If the case (c) holds for some pair $\{i_c,j_c\}$, then
neither $i_c$ nor $j_c$ can be involved in a case (a).
\\
{\it Proof.}
Assume $i_c=i_a$. Then, since $i_a$ has only one neighbor
we have $\hat{\NN}(i_c) = \emptyset = \hat{\NN}(j_c)$ which
is not allowed [see also (i)].
If $i_c=j_a$ then $i_a \in \NN(i_c).$
We cannot have $i_a = j_c$ [see (i)] but we can also not have
$i_a \in \NN(j_c)$, since $i_a$ has only one connection. This
is then a contradiction to $\hat{\NN}(i_c) = \hat{\NN}(j_c).$

\item[(iii)]
If the case (c) holds for some pair $\{i_c,j_c\}$, then neither $i_c$ nor $j_c$
can be involved in a case (b).
\\
{\it Proof.}
Assume $i_c=i_b$, then $j_c \neq j_b$ since property
(i) holds. We have that $j_b \notin \NN(i_b) = \NN(i_c)$
which implies that $j_b \notin \NN(j_c).$ On the other hand,
we have that $j_c \in \NN(i_c)=\NN(i_b)\stackrel{(b)}{=}\NN(j_b).$
This means that both $i_b$ and $j_b$ are connected to $j_c,$
hence $j_b \in \NN(j_c),$ which is a contradiction.

\item[(iv)]
If the case (b) holds for some pair $\{i_b,j_b\}$ then there
exists no pair $\{i_a,j_a\}$, for which the case (a) holds,
such that $i_b=j_a.$ But there might exist a pair $\{i_a,j_a\}$,
of case (a) such that $i_b=i_a$.
\\
{\it Proof.}
Assume that $i_b=j_a$. Then $i_a \in \NN(i_b),$
but  $i_a \in \NN(j_b)$ cannot be true since  $i_a$
has only one connection to the rest of the graph.
So $\NN(i_b) \neq \NN(j_b)$ which is in contradiction
to the case (b). The second part of the property is
proved by providing an example. For this, one can
just take a linear graph with three vertices, numerated
from left to right. Then, the first and the second qubits
are $i_a$ and $j_a$ respectively, whereas the first and
the third qubits are $i_b$ and $j_b$.

\item[(v)]
For a pair $\{i_a,j_a\}$ of case (a), the qubit $i_a$
may also appear in a case (b). The qubit $j_a$ cannot
appear in a case (b) or (c), but it may also appear in another
case (a) pair $\{\tilde{i_a},\tilde{j_a}\}$ but only as
$\tilde{j_a}.$
\\
{\it Proof.}
The first part just follows from the previous points. For the last
part, one may consider the example in (v), there, the second (middle)
qubit takes part in two different cases (a). It is clear, that
$j_a$ cannot be $\tilde{i_a}$ for some other case (a), unless we
have a two-qubit graph.

\item[(vi)]
For a pair $\{i_b,j_b\}$, both $i_b$ and $j_b$ can
take part in another pair of case (b). In analogy,
for a pair $\{i_c,j_c\}$, both $i_c$ and $j_c$ can
take part in another pair of case (c).
\\
{\it Proof.}
We just give examples where this may happen. For case (b),
consider a graph where one central qubit has a single
connection to $m\ge 2$ other qubits. For case (c) consider
$m\ge 3$ qubits which are all interconnected.

\end{itemize}

These are all interesting cases we have to consider. In summary,
this proves the following: A single qubit $k$ can only take part
in two different cases $\{i,j\}$ and $\{m,n\}$, if the nontrivial
Pauli matrix contributions from $\vr_{ij}$ and $\vr_{mn}$ in
Eqs.~(\ref{casea}, \ref{caseb}, \ref{casec}) have the same Pauli
matrix on the qubit $k.$ For instance, in property (iv), if
$k=i_a=i_b$, then in Eqs.~(\ref{casea}, \ref{caseb}) both Pauli
matrices on qubit $j$ have are the same, namely $X_k.$ The same is
true for the property (v), here, when $k=j_a=\tilde{j_a},$ the
Pauli matrix is $Z_k.$ If (vi) occurs,
then the  observable will be $X_k$ in case (b) and $Y_k$ in case (c).

This implies that for an arbitrary graph with more than three
qubits, we can choose a fixed observable $O^{(k)} \in \{X_k, Y_k, Z_k\}$
for any qubit $k$, such that any non-vanishing
two-body correlation can be represented by a tensor product of the
$O^{(k)}.$ Note that since just a single observable $O^{(k)}$
occurs on each qubit, it can be changed to $Z_k$ with a
local unitary operation, which does not influence the entanglement
properties of the graph state

{\it Part III ---}
Finally, we can prove the claim of Theorem 10. Consider the observable
\be
\MM_G = \bigotimes_{k=1}^{N} O^{(k)},
\ee
which is constructed for a special graph state $\ket{G}$.

This observable has $2^N$ eigenvectors (which are product
vectors) and measuring this observable results in $2^N$
probabilities, which can be labeled by the possible results
$\pm 1$ for the $N$ qubits. In order to do this, we
introduce a vector $\vec \alpha$
where the entry $\alpha_k$ is equal to $0$ if the result on qubit $k$
is $+1$ and equal to $1$ is the result on qubit $k$ is $-1$.
We denote the eigenvectors by $\ket{\vec\alpha}$.
Note that they are product vectors $\ket{\vec\alpha}=\otimes_k\ket{\alpha_k}$
defined by $O^{(k)}\ket{\alpha_k}=(-1)^{\alpha_k}\ket{\alpha_k}$.

With this notation, we can write the expectation value of the observable
${\MM_G}$ as $\mean{\MM_G}=\sum_{\vec\alpha}(-1)^{\sum_k \alpha_k} p_{\vec\alpha}$,
where $p_{\vec\alpha}=|\braket{\vec\alpha}{G}|^2$.
Further, from the measurement probabilities $p_{\vec\alpha}$,
it is possible to compute {\it all} non-vanishing two-point correlations of the graph state $\ket{G}
$, since they are, as mentioned above, of the type $\mean{O^{(i)}O^{(j)}}$, where both $O^{(i)}$ and
$O^{(j)}$ appear in the observable $\MM_G$.

However, the fully separable state
$\vr_{\rm fs}=\sum_{\vec\alpha} p_{\vec\alpha}\proj{\vec\alpha}$ with
exactly the same probabilities $p_{\vec\alpha}$ will reproduce all the non-vanishing
two-body correlations of the graph state. In fact, it also reproduces the vanishing ones:
Let us assume for simplicity that $\MM_G=Z^{\otimes N}.$ Then, the state $\vr_{\rm fs}$
can be chosen such that it will reproduce all correlations (also vanishing two-body
correlations or the non-vanishing higher order correlations) of the graph state in
the basis defined by $\MM_G=Z^{\otimes N}.$ Other two-body correlations, such as
$\mean{X_i Y_j}$, will vanish for the graph state. However, we can expand $\vr_{\rm fs}$
in terms of tensor products of Pauli matrices as
\be
\vr_{\rm fs} = \sum_{k_i=0,1} \beta_{k_1 ... k_N} Z^{k_1} \otimes Z^{k_2} \otimes ... \otimes Z^{k_N}
\ee
since $\ketbra{\alpha}=(1/2^N) \bigotimes_{k=1}^N [\eins + (-1)^{\alpha_k}Z_k]$ can be expressed in this form.

This implies, that for $\vr_{\rm fs}$ the two-body correlations (and also higher correlations)
in a different basis than $\MM_G=Z^{\otimes N}$ vanish.
All in all, the fully separable state $\vr_{\rm fs}$ has the same
one- and two-qubit correlations than the graph state, so the reduced
density matrices must be the same.
\qed

\section{Conclusion}
\label{disc5}
In conclusion, we have derived a generalization of the covariance
matrix criterion to the multipartite case. We have seen that the
resulting criterion is strong, but some highly entangled states are
not detected. This is, however, an example of a more general phenomenon,
which is that no criterion based on two-particle correlations can
detect graph state entanglement.

There are several ways to extend our results in the future. First, one
can aim to develop a criterion based on covariance matrices  for the
detection of genuine multipartite entanglement. The present criterion
can only exclude full separability. Second, one can try to connect the
entanglement parameter $t$ in the formulation of the multipartite CMC in
Eq.~(\ref{sdph}) with multiparticle entanglement measures, similarly as it
has been done in for the bipartite CMC \cite{cmquanti}. Finally, it would be
interesting to gain further insight into the question which types of
entanglement can be verified by two-particle correlations and which not.
This  question is also experimentally relevant, since in some experiments
only two-particle correlations can be measured.

\subsection*{Acknowledgements}
We thank J. Eisert, B. Jungnitsch, M. Kleinmann, C. Knapp, S. Niekamp, N. L{\"u}tkenhaus and G. T{\'o}th for discussions and
acknowledge support from the FWF (START prize and SFB FOQUS) and the EU (NAMEQUAM and QICS). This work was finished after O.G. moved to the IQC in Waterloo, and he is grateful for support from the Industry Canada and NSERC Strategic Project Grant (SPG) FREQUENCY.\\[.4cm]


\setcounter{thm}{0} \setcounter{equation}{0} \setcounter{figure}{0}
\renewcommand{\theequation}{A-\arabic{equation}}
\renewcommand{\thefigure}{A-\arabic{figure}}
\renewcommand{\thethm}{A-\arabic{thm}}

\section*{{\bf APPENDIX}}
\label{append}
{\it Proof of Lemma \ref{lemma3part}:}
We shall split the proof into four parts:

(i) If all off-diagonal elements of
$\eta$ are positive, the claim follows immediately.

(ii) If exactly two
off-diagonal elements of $\eta$, say, $d$ and $f$
are negative then, we have that $\tilde{\eta} = D \eta D^T$
with a diagonal matrix $D={\rm diag}(1, -1, 1),$ so the positivity
of $\eta$ is equal to the positivity of $\tilde{\eta}.$

(iii) If only one element is negative, say $e<0$, we have to prove
that positivity of $\eta(e)$ guarantees the positivity of
$\tilde{\eta}=\eta(|e|)$. To this end we show an equivalence of the
positive semidefiniteness of any $3\times 3$ matrix $A$ and
positive semidefiniteness of any of its $2\times 2$ submatrix
and its determinant. It is clear that the implied conditions on the submatrices and the determinant
are necessary.
Sufficiency is proved considering the characteristic
polynomial of a $3\times 3$ matrix $A$, which is given by
\be
\chi_A(\lambda)={-\lambda^3+Tr(A)\lambda^2-\sum_{i=1}^3\det(A_{ii})\lambda}+
\det(A).
\label{charpol}
\ee
Here, $A_{ii}$ is the $2\times 2$ submatrix of $A$, which arises if the $i$-th
column and row are deleted.

Since $A$ is symmetric, it has real eigenvalues and all roots of this
polynomial are real. If $\det(A_{ii})\geq 0$ and $\det(A)\geq 0$, then
the zeroes clearly have to be nonnegative, because then for $\lambda < 0$
one has  $\chi_A(\lambda)>0.$

So let us consider
\bea
\det(\eta) &= abc + 2e|f||d| - be^2 - a|f|^2 - c|d|^2 \nonumber\\
&= abc - 2|e||f||d| - be^2 - a|f|^2 - c|d|^2\geq 0.
\eea
{From} this it follows that $\det(\tilde{\eta})=\det(\eta)+ 4|e||f||d| \geq \det(\eta) \geq 0,$
moreover, the determinants of $2\times2$-submatrices of $\tilde{\eta}$ are the same as the ones from
$\eta$ and hence positive. Therefore, $\tilde{\eta}$ must be positive semidefinite.

(iv) Finally, if all off-diagonal elements of
$\eta$ are negative we can first flip the sign of two of
them as in (ii), and then we arrive at the third case.
\qed


\end{document}